\documentclass{iaus}
\usepackage{graphics}
\input{psfig}
  \checkfont{eurm10}
  \iffontfound
    \IfFileExists{upmath.sty}
      {\typeout{^^JFound AMS Euler Roman fonts on the system,
                   using the 'upmath' package.^^J}%
       \usepackage{upmath}}
      {\typeout{^^JFound AMS Euler Roman fonts on the system, but you
                   dont seem to have the}%
       \typeout{'upmath' package installed. iaus.cls can take advantage
                 of these fonts,^^Jif you use 'upmath' package.^^J}%
      }
  \else
  \fi

  \checkfont{msam10}
  \iffontfound
    \IfFileExists{amssymb.sty}
      {\typeout{^^JFound AMS Symbol fonts on the system, using the
                'amssymb' package.^^J}%
       \usepackage{amssymb}%

      }{}
  \fi

  \IfFileExists{amsbsy.sty}
    {\typeout{^^JFound the 'amsbsy' package on the system, using it.^^J}%
     \usepackage{amsbsy}}
    {}

\newsavebox{\astrutbox}
\sbox{\astrutbox}{\rule[-5pt]{0pt}{20pt}}

\def\HI{H{\,\small I}}
\newcommand{\ltsima} {$\; \buildrel < \over \sim \;$}
\newcommand{\gtsima} {$\; \buildrel > \over \sim \;$}
\newcommand{\lta} {\lower.5ex\hbox{\ltsima}}
\newcommand{\gta} {\lower.5ex\hbox{\gtsima}}

\newcommand{\msun}{{$M_\odot$}}

\title[Radio-activity and ISM in radio galaxies]{The interplay between 
radio-activity and the ISM in radio galaxies}

\author[R. Morganti]
{R. Morganti$^1$}

\affiliation{$^1$Netherlands Foundation for Research in Astronomy, Postbus 2,
NL-7990 AA, Dwingeloo, NL
email: morganti@astron.nl}

\pubyear{2004}
\volume{222}
\pagerange{1--8}
\date{?? and in revised form ??}
\jname{The Interplay among Black Holes, Stars and ISM \\in Galactic Nuclei}
\editors{Th. Storchi Bergmann, L.C. Ho \& H.R. Schmitt, eds.}
\begin{document}

\maketitle

\begin{abstract}

Radio-loud AGNs can inhabit regions with a very rich ISM.  The presence of
this rich medium is likely related to the origin and evolution of the host
galaxy and of the active nucleus. Recent observations show that a large
fraction of radio galaxies contains a significant young stellar
population. This supports the idea that mergers are responsible for both the
starburst phase and the triggering of the nuclear activity. The gas that
reaches the central regions can have quite disturbed kinematics, likely due to
the effects of the AGN activity and in particular of the powerful radio jets.
The recent detection of fast nuclear gas outflows, observed both in ionised
and neutral gas, is giving new and important insights into the physical
conditions of the gaseous medium around the nucleus and the interaction
between the AGN and this medium.  Finally, as another example of the interplay
between the radio activity and the ISM, the possibility of star formation
induced by the passage of the radio jet will be discussed.

\end{abstract}

\firstsection 

\section{Introduction}

This review concentrates on one particular type of active nucleus, radio
galaxies, where the presence of large-scale and powerful jets makes it
particularly interesting to look for the importance of the interplay between
these structures and the interstellar medium (ISM).  Radio galaxies are
invariably associated with early-type host galaxies, i.e.\ galaxies that were
thought until not so long ago to have a relatively uninteresting ISM.  It is clear
now that at least a fraction of them has a very rich ISM likely related to
their origin.

The origin of activity in galaxies is often explained as triggered by merger
and/or interaction processes.  Torques and shocks during the merger can remove
angular momentum from the gas in the merging galaxies and this provides
injection of substantial amounts of gas/dust into the central nuclear regions
(see e.g.\ Mihos \& Hernquist 1996). This also appears to be the case for
radio galaxies as suggested by morphological and kinematical evidence (e.g.\
Heckman et al.\ 1986, Tadhunter et al.\ 1989). On the other hand, the phase of
nuclear activity triggered in this way is now increasingly recognised to play
an important role in the evolution of the galaxy itself.  Particularly
important in this respect are gas outflows that can be generated by this
activity and the effect they can have on the interstellar medium (ISM). This
feedback can be extremely important for the evolution of the galaxy, up to the
point that it could limit the growth of the nuclear black-hole (e.g.\ Silk \&
Rees 1998, Wyithe \& Loeb 2003).  Thus, the processes of assembly of the host
galaxy, the supply of gas to the central region, as well as the effects that
the triggering of the (radio) activity has on this gas, are tightly related
and essential for our understanding of radio galaxies.

Here I will review some recent results related to these topics and in
particular I will discuss two possible effects of the interplay between radio
activity and ISM: gaseous outflows and jet induced star formation.

\section{Starburst, AGN phase and the order of events}

The origin of radio galaxies can be studied using the stellar population of the
host galaxy. This gives information on the age of the last merger as well as
the type of merger that triggered the activity.  Complementary to this, is the
study of the large-scale \HI\ that can be used as another tracer of the
formation process of their host galaxy.

\subsection{Results from the stellar population}

A number of surveys have shown -- through the modelling of spectroscopic and
polarimetric data -- that young stellar populations (YSP) make a significant
contribution to the optical/UV continua in 25 - 40\% of powerful radio galaxies
at low and intermediate redshift ($z<0.7$, Aretxaga et al.\ 2002, Tadhunter et
al.\ 2002, Wills et al.\ 2002, 2004). Similar results have been obtained also
from {\it UV} imaging (Allen et al.\ 2002).

Furthermore, the study of a few selected radio galaxies (3C~293, 3C~305 and
4C12.50, Tadhunter et al.\ 2004) has shown that the YSP are relatively old (0.1
- 2 Gyr), massive ($10^9 < {M}_{\rm YSP} < 5\times 10^{10}$ $M_{\odot}$) and
make up a large proportion ($\sim 1 - 50$\%) of the stellar content.  These
results are consistent with the idea that AGN activity, at least in some radio
galaxies, is triggered by major gas-rich mergers and confirm that an
evolutionary link exists between radio galaxies and luminous- and
ultra-luminous infrared galaxies (Tadhunter et al.\ 2004) as suggested earlier
e.g., by the study of the molecular gas (see e.g.\ Evans et al.\ 1999).  These
mergers will trigger both the starburst and the AGN activity.  Interestingly,
these results also suggest that the radio activity is triggered relatively late
in the merger sequence. This delay could be due to the fact that the starburst
phase momentarily clears the central regions from gas and that it takes some
time before other ``fuel'' can concentrate again in the nuclear regions and
trigger, this time, the AGN. However, the fact that a large fraction of radio
galaxies do not show a young stellar population component also indicates (as
previously suggested e.g.\ Heckman et al.\ 1986) that radio sources are not
triggered by a single type of merger or at a particular stage of the merger
sequence (Wills et al.\ 2002).

\subsection{Results from \HI\ studies}

A further way to investigate the origin and evolution of the host galaxy is by
studying the large scale neutral hydrogen.  As already found in ``normal''
early-type galaxies (see e.g.\ Oosterloo et al.\ 2004), also some radio galaxies
show very large (up to 200 kpc in size and masses up to $10^{10}$ $M_{\odot}$)
and regular disks (of low column density) neutral hydrogen (see e.g.\ the case
of PKS~B1718-649, V\'eron-Cetty et al.\  1995 and B2~0648+27, Morganti et
al.\ 2003a). A systematic survey (Emonts et al.\ in prep.) is now looking for
more cases like these.  One example, the large \HI\ disk detected around the
compact radio galaxy B2~0258+35, is shown in Fig.\ 1. Given their size and
regular appearance, these disks must be quite old ($> 10^9$ yr), hence they
are not related to recent accretions, in agreement with what is found from the
study of the stellar population.  Thus, also from the study of the \HI\ we
derive that the radio activity starts late after the merger.  Interestingly,
all these large and rich \HI\ disks appear to be, so far, associated with
compact (steep spectrum) radio sources, typically considered young radio
galaxies (O'Dea 1998).

\begin{figure*}
\centerline{\psfig{figure=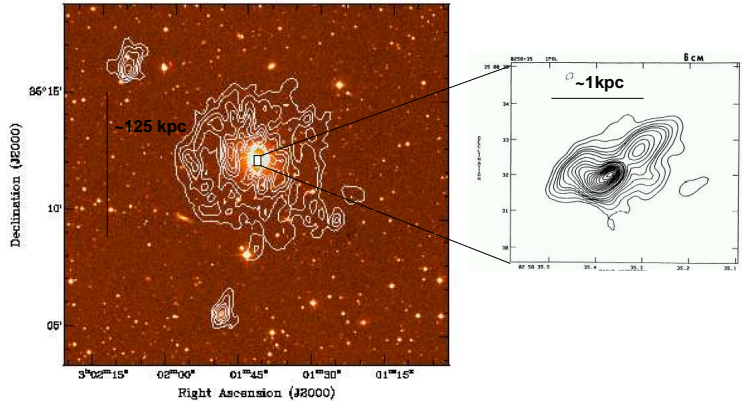,angle=0,width=10cm}
}
\caption{{\sl Left} \HI\ total intensity (contours) superimposed to an optical 
image of B2~0258+35 (from Emonts et al. in prep.). {\sl Right} The
continuum image (from Fanti et al. 1986).  }
\end{figure*}

\section{Outflows of neutral and ionized gas in the central regions}

Given the origin described above of, at least some, radio galaxies, one can
expect large amounts of gas to be transported and concentrated in the central
regions.  Examples of high nuclear concentration of gas, in particular
molecular gas, are indeed known (e.g.\ Evans et al.\ 1999, 2004). The ionised
and neutral gas observed in the central regions of radio galaxies is often
associated with circum-nuclear disks, as already discussed in other
contributions in these Proceedings. However, gas is not only observed in these
``settled'' structures.  In young (i.e.\ CSS and GPS, O'Dea 1998) or recently
re-started radio galaxies, a cocoon of gas may still surround the active
nucleus and the effects produced on it by the AGN activity can be observed. For
example, the radio jets are expected to go through some ``struggle'' to be able
to emerge from the nuclear regions of these young radio sources.  Indeed, broad
optical (forbidden) emission lines are typically observed in CSS/GPS sources,
indicating the presence of gas with disturbed kinematics as a result of an
interaction of the radio plasma with the ISM (Gelderman \& Whittle 1994, O'Dea
et al.\ 2002).

In addition to this, gaseous outflows are now unambiguously detected in a
number of young radio sources.  The two best examples are PKS~1549--79 and
4C~12.50 (Tadhunter et al.\ 2001, Holt et al.\ 2003). In the case of 4C~12.50,
complex, multi-component emission line profiles are observed at the position
of the nucleus. The broadest component has FWHM$\sim 2000$ km/s, is blueshifted
by $\sim 2000$ km/s\ with respect to the halo of the galaxy, has a large
reddening and high density ($n_{\rm e} >$ 4200 cm$^{-3}$). This component is
therefore interpreted as the material closer to the radio jet and interacting
with it. This interaction produces the observed outflow.  A more quiescent (and
extended) component of ionized gas is also observed and it is associated with
the cocoon of gas surrounding the radio source but not yet interacting with
it. Thus, 4C~12.50, as well as PKS~1549-79, appears to be a young radio galaxy
with nuclear regions that are still enshrouded in a dense cocoon of gas and
dust. The radio jets are now expanding through this cocoon, sweeping material
out of the nuclear regions (Holt et al.\ 2003).  Interestingly, high resolution
(VLBI) observations of the \HI\ in 4C12.50 show that  deep and relatively
narrow \HI\ absorption (observed at the systemic velocity) is indeed associated
with an off-nuclear cloud ($\sim 50$ to 100 pc from the radio core) with a
column density of $\sim 10^{22}\ T_{\rm spin}/(100\ {\rm K}$) cm$^{-2}$ and an
\HI\ mass of a few times $10^5$ to $10^6$
\msun. Although this gas will not be able to confine the radio source,
it could, however, be able to {\sl momentarily} destroy the path of the jet as
shown also by numerical simulations (Bicknell et al.\ 2003). Thus, this
interaction can influence the growth of the radio source until the radio plasma
clears its way out.  It is interesting to point out the fact that all cases of
fast outflows (both of ionised and neutral gas, see below) have been observed
so far in radio galaxies with YSP. Thus, this could indicate that the galaxy is
indeed in a particular stage of its evolution where large amounts of gas are
still present in the central region.  The hollowed out bi-conical structures
observed, e.g., in the nearby radio galaxy Cygnus~A (Tadhunter et al.\ 1999),
might represent the aftermath of this process once the source reaches a later
stage of the evolution.

\subsection{Fast outflows of neutral hydrogen}

Extremely intriguing is the discovery of a number of radio galaxies where the
presence of fast outflows is associated also with {\sl neutral} gas. The best
example of this phenomenon found so far is the radio galaxy 3C~293 (Morganti et
al.\ 2003b) but more cases have now been observed (see also Oosterloo et
al. these Proceedings).  Fig.~2 shows the case of 4C~12.50. The broad and
shallow (with optical depth typically only a fraction of a \%) \HI\ absorption
features detected against the central regions have a full width at zero
intensity ranging between 800 and 2000 km/s, mostly blueshifted relative to the
systemic velocity.  The broad \HI\ absorption represents a fast outflow of {\sl
neutral} gas from the central regions of these radio galaxies.  Interestingly,
in the few cases known so far, the fast outflows detected in neutral and ionised
gas show similar kinematics (Oosterloo et al.\ these Proceedings). This suggests
that both are due to the same mechanism.


\begin{figure*}
\centerline{
\psfig{figure=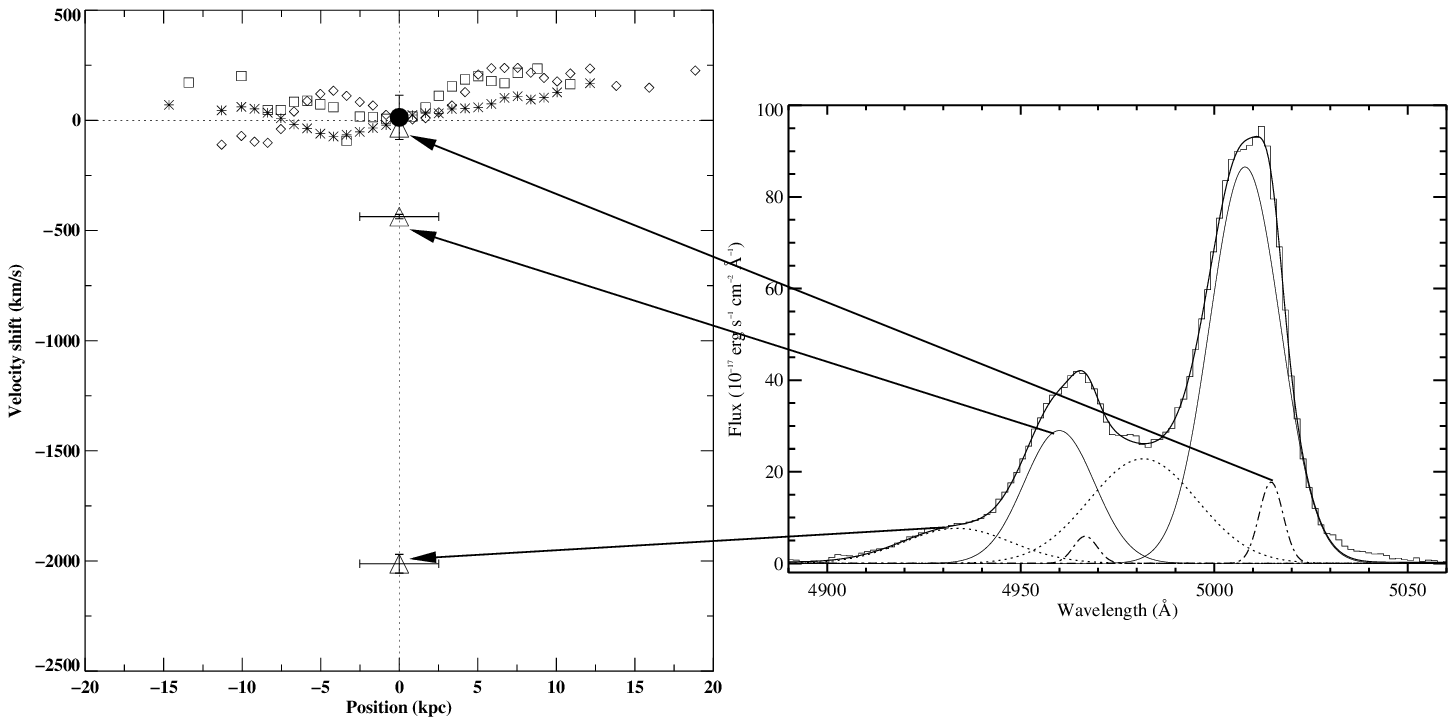,angle=0,width=8cm}
\psfig{figure=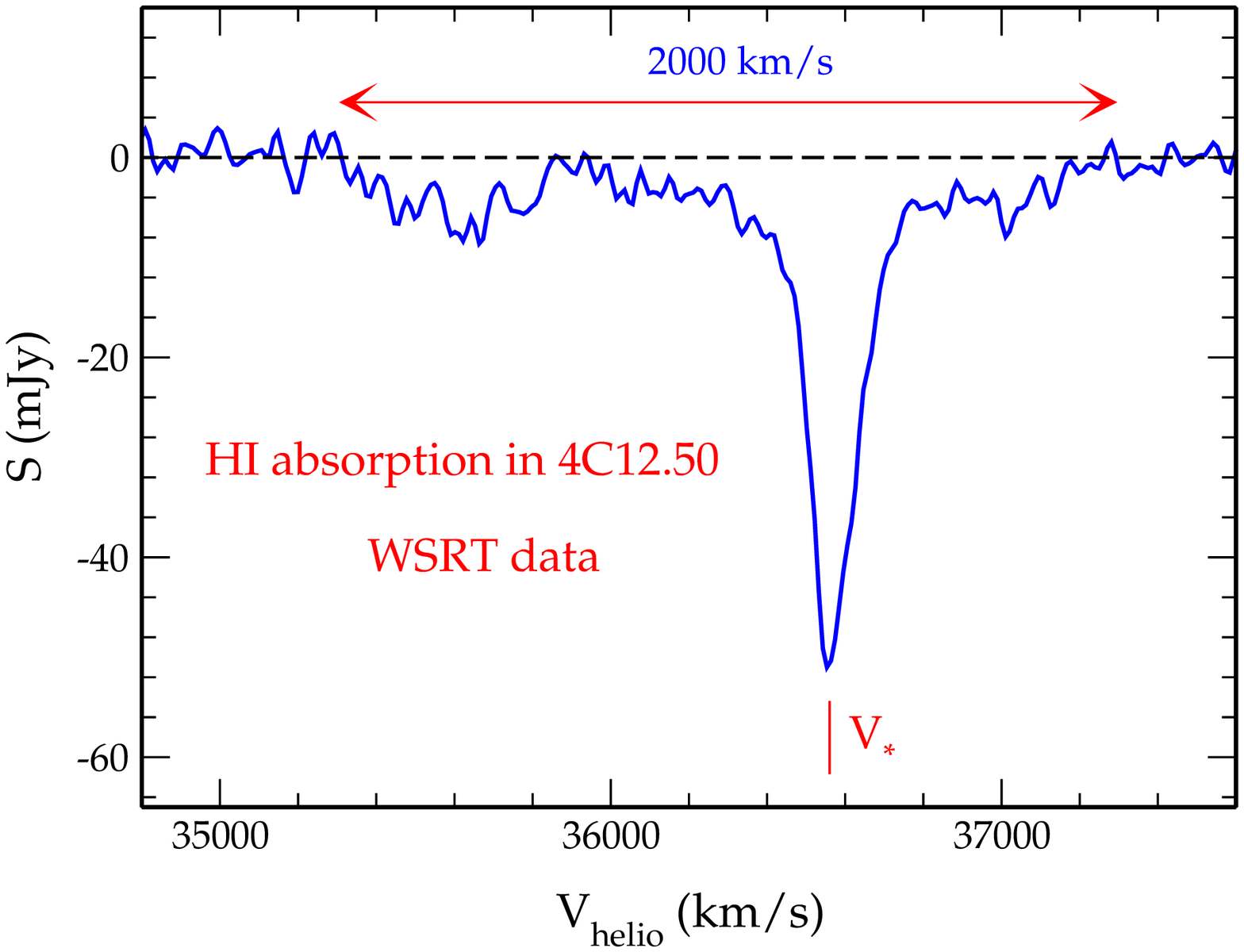,angle=0,width=6cm}}
\caption{{\sl Left} Radial velocity profile of the extended gaseous
halo of 4C~12.50 (from Holt et al. 2003). Key: $\ast$, $\Diamond$ and $\Box$:
highly extended {[O II]} emission (narrowest component); $\bullet$: \HI\ 21-cm
absorption; $\triangle$: 3 components of {[O III]} in the nucleus.  The [OIII]
profile with the components from the fit is also shown.  {\sl Right} The \HI\
profile (from the WSRT) of 4C~12.50 showing the broad absorption. The spectra are plotted in flux density (mJy) against optical
heliocentric velocity in km/s.}
\end{figure*}

Gas outflows of ionised gas are detected in other type of AGNs (such as quasars
and Seyfert galaxies, see e.g.\ Kriss these proceeding). These outflows can be
produced in regions close to the accretion-disk of the AGN (Elvis 2000), or the
gas could be accelerated through radiation and/or wind pressure originating in
dusty narrow-line regions (Grove et al.\ these Proceedings) or starburst winds
(Heckman et al.\  1990). However, for radio-loud objects gas outflows can be due
to the interaction between the radio plasma and the ISM medium. To discriminate
between all these mechanisms as origin for the outflows in radio galaxies it is
necessary to identify the location where the outflow (both in ionised and
neutral gas) is occurring in relation to the radio structure. So far this has
been done only in the case of 3C~293 (see Oosterloo et al.\ these Proceedings)
where the outflow appears to be located at the position of a radio lobe.  A
possible model of what is happening is that the radio plasma jet hits a
(molecular) cloud in the ISM.  As a consequence of this interaction, part of
the gas is ionised and its kinematics is disturbed by the shock.  Once the
shock has passed, part of the gas may have the chance to recombine and become
neutral, while it is moving at high velocities.  A similar explanation has been
proposed for other objects (see e.g.\ Conway \& Schilizzi 2002).  As result of
such an interaction, star formation might also occur (see below).  Finally, it is
worth to point out that these cases of extreme kinematics of the gas indicate
once again that (as pointed out in more detail in other talks at this
conference) one has to be very careful in using either the neutral or the
ionised gas to derive black hole masses.

\begin{figure*}
\centerline{
\psfig{figure=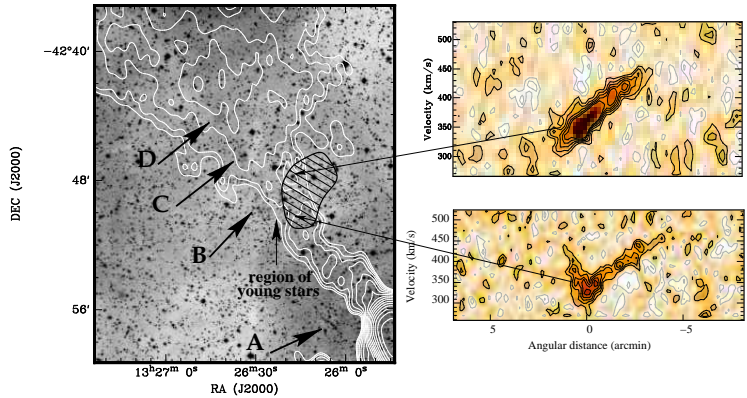,angle=0,width=10cm}}
\caption{{\sl Left} The radio continuum (contours) superimposed to an optical
image of Centaurus~A (from Morganti et al. 1999, the regions of the optical
filaments are labelled). The region with \HI\ emission is shaded.  {\sl Right}
two position-velocity diagrams showing (top) the smooth, regular velocity
pattern observed in most of the cloud and (bottom) the region where anomalous
velocities are observed.}
\end{figure*}

\section{Jet-induced star formation}

Another effect of the interplay between the AGN activity and the ISM is star
formation induced by the passage of a radio jet. This is particularly relevant
as it has been considered as a possible mechanism to explain the UV continuum
emission observed in the host galaxies of high-$z$ radio galaxies and the
"alignment effect" between radio emission and this continuum (Rees 1989).

Recent numerical simulations have shown that the interaction of a radio jet
with a clumpy medium can produce fragmented clouds that can quickly cool and
condense (see Mellema et al.\ 2002, Fragile et al.\ 2004). Thus, the interaction
and the consequent production of shocks from the radio jet (even with moderate
velocity) has, indeed, the potential to trigger large-scale star formation in a
galaxy, (Fragile et al.\ 2004).  Few cases are known in nearby radio galaxies
(see e.g.\ van Breugel et al.\ 2004) and one of the best example is observed in
the radio galaxy Centaurus~A.  The detection of groups of young stars in the
northern lobe of this galaxy and close to the location of the radio jet, has
been explained by different authors as due to gas shocked by the passage of
this jet (see e.g.\ Graham 1998). Next to this region ($\sim 15$ kpc from the
centre) an \HI\ cloud has also been observed (Schiminovich et al.\ 1994). This
cloud has been interpreted as being  part of a single inclined ring
regularly rotating in a circular orbit.  However, observations of this cloud
with higher velocity resolution (Oosterloo
\& Morganti 2004 in prep.) reveal anomalous velocities of the \HI.
The position-velocity diagram shown in Fig.~3 shows that an abrupt change in
velocity of about 100 km/s/kpc happens in the southern tip of the cloud,
i.e.\ the region closer (in projection) to the passage of the radio jet. Such
change in velocity can only be explained as results of the interaction between
the jet and the \HI\ cloud. This is interpreted as  further evidence that, at
that location, an interaction is occurring between the radio jet and the
ISM. This gives further evidence that the young stars observed in that region
are the result of such an interaction and that the neutral hydrogen provides the
reservoir that makes the star formation possible.

\begin{acknowledgements}
I would like to thank the organisers, in particular Thaisa
Storchi Bergmann, for inviting me and for the financial support.  I am
also very grateful to my collaborators, in particular Tom Oosterloo
and Clive Tadhunter, with whom I have done a significant fraction of
the work discussed here.

\end{acknowledgements}

\end{document}